\documentclass[fleqn,10pt]{wlscirep}
\usepackage[utf8]{inputenc}
\usepackage[T1]{fontenc}
\usepackage{xcolor}
\usepackage{footnote}
\usepackage[switch]{lineno}
\usepackage{amsmath,amsfonts}
\usepackage{algorithmic}
\usepackage{algorithm}
\usepackage{array}
\usepackage{textcomp}
\usepackage{stfloats}
\usepackage{url}
\usepackage{verbatim}
\usepackage{graphicx}
\usepackage{cite}
\usepackage{hyperref}

\title{Predicting Nodal Influence via Local Iterative Metrics}

\author[1]{Shilun Zhang}
\author[1]{Alan Hanjalic}
\author[1,*]{Huijuan Wang}
\affil[1]{Faculty of Electrical Engineering, Mathematics, and Computer Science, Delft University of Technology, Mekelweg 4, 2628 CD, Delft, The Netherlands}
\affil[*]{H.Wang@tudelft.nl}

\begin{abstract}
Nodal spreading influence is the capability of a node to activate the rest of the network when it is the seed of spreading. Combining nodal properties (centrality metrics) derived from local and global topological information respectively is shown to better predict nodal influence than a single metric. In this work, we investigate to what extent local and global topological information around a node contributes to the prediction of nodal influence and whether relatively local information is sufficient for the prediction. We show that by leveraging the iterative process used to derives a classical nodal centrality such as eigenvector centrality, we can define an iterative metric set that progressively incorporates more global information around the node. We propose to predict nodal influence using an iterative metric set that consists of an iterative metric from order $1$ to $K$ that are produced in an iterative process, encoding gradually more global information as $K$ increases. Three iterative metrics are considered, which converge to three classical node centrality metrics respectively. Our results show that for each of the three iterative metrics, the prediction quality is close to optimal when the metric of relatively low orders ($K\sim4$) are included and increases only marginally when further increasing $K$. The best performing iterative metric set shows comparable prediction quality to the benchmark that combines seven centrality metrics, in both real-world networks and synthetic networks with community structures. Our findings are further explained via the correlation between an iterative metric and nodal influence, the convergence of iterative metrics and network properties. 
\end{abstract}
\begin{document}
\maketitle
\thispagestyle{empty}
\noindent
\section*{Introduction}
Spreading processes are ubiquitous in various systems of nature and society. Examples include the spreading of epidemics, the propagation of information, and cascade of failures. Complex networks, usually considered as the underlying structure of such systems, provide the substrate upon which the spreading process unfolds via links connecting nodes. The spreading influence of a node represents the extent to which the node, where the spread originates, can eventually activate other nodes in the network. For a given spreading process, the spreading influence of a node is defined as the expected outbreak size when the spreading process starts from the node, also called the seed node. Due to the topological heterogeneity of nodes in many real networks \cite{newman2018networks}, some nodes may have significantly higher spreading influence and are evidently more influential than the other nodes \cite{lloyd2005superspreading, pastor2015epidemic, hu2018local}. Identifying these influential nodes and predicting their spreading influence is crucial for controlling the spread of epidemics \cite{woolhouse1997heterogeneities, pei2013spreading} or rumors \cite{chen2020rumor, bovet2019influence}, promoting strategic marketing \cite{watts2007influentials, leskovec2007dynamics, kempe2005influential}, quantifying the impact of researchers and publications \cite{zhou2012quantifying}, and more \cite{zhan2020susceptible, wang2021local, zhang2019long}.

Two generic influence prediction problems have been addressed in prior research. The first involves identifying the most influential nodes among all nodes based on the given network topology. To solve this problem, previous studies have proposed to rank nodes by a single nodal topological metric, so-called centrality metric \cite{kitsak2010identification,lu2016vital,li2015correlation}, which encodes either local \cite{chen2012identifying, lawyer2015understanding} or global \cite{kitsak2010identification,klemm2012measure} topological information around a given node.
The highest-ranked nodes are then identified as the most influential ones. Nonetheless, these prior work suggests that no single centrality metric can outperform all other centralities for different epidemic parameters and in diverse types of networks, since a centrality metric only captures a certain topological feature of a
 node. It has been shown that nodal degree, i.e., number of $1$-hop neighbors, is more (less) predictive than eigenvector centrality \cite{maharani2014degree} when the spreading rate is small (large) \cite{liu2016locating, pei2013spreading}. The coreness better predicts the top spreaders than nodal degree in Susceptible-Infected-Recovered model below epidemic threshold. Further studies put forward methods to integrate local and global centralities or their rankings. Zhe Li et al. \cite{li2022identifying} used the sum of normalized degree, eigenvector centrality, and coreness as the mass of a node in a gravity model to derive a new nodal metric. Andrea Madotto et al. \cite{madotto2016super} aggregated the ranking lists by local and global node centralities to produce a new ranking list based on the correlations between the rankings. These methods usually exhibit better performance than merely using a local or global centrality. 

In many practical scenarios, it is possible to observe or derive the spreading influences of a small fraction of nodes. For example, the average number of retweets of content posted by a node can be used as an approximation of the spreading influence of the node \cite{pei2013spreading, pei2014searching}. This motivates the second influence prediction problem: identify the most influential nodes given the network topology and the influence of a small fraction of nodes. Bucur \cite{bucur2020top} recently proposed to train a statistical model on the set of nodes whose spreading influences are known to classify the rest nodes into binary classes, representing whether a node is among the top (e.g., top $10\%$) influential ones or not. The statistical model maps the relation between the class of a node in spreading influence and centrality metrics including both local centrality metrics like degree and global centrality metrics like betweenness \cite{wang2008betweenness} and eigenvector centrality. These centrality metrics were shown to be able to complement each other to achieve universally good performance in locating the most influential nodes across various real-world networks. However, global centrality metrics have a high computational complexity, which limits their application to large-scale networks. Moreover, the non-trivial correlation among different metrics makes it difficult to interpret to what extent global nodal properties are needed to estimate nodal spreading influence.

\begin{figure*}[ht]
\centering
\includegraphics[scale=0.65]{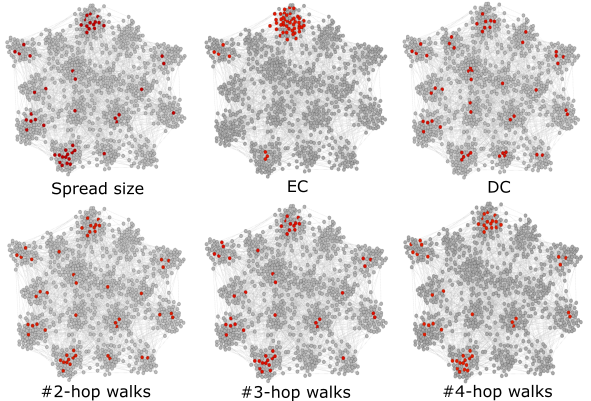}
\caption{Location of top ranked nodes in a network generated by LFR model. The red-colored nodes are the top $10\%$ nodes when nodes are ranked by spread size (top left), eigenvector centrality (EC, top middle), degree centrality (DC, top right), 2-hop walk counts (bottom left), 3-hop walk counts (bottom middle), and 4-hop walk counts (bottom right), respectively.}
\label{fig:illustration}
\end{figure*}

To bridge this gap, we will systematically explore two foundational questions: how local and global topological information around a node contribute to the prediction of nodal spreading influence, and how to predict nodal spreading influence efficiently using relatively local information. The general prediction task is considered: given the topology of a network and the spreading influences of a fraction of nodes how to predict the spreading influences of the other nodes in the network, beyond their ranking. To solve the prediction task, a node-level regression model is trained on the set of nodes whose spreading influences are known and used to predict the influences of the remaining nodes. To understand how local and global topological information contribute to the prediction, we design the input of the regression model based on nodal properties as follows. We show that by leveraging the iterative process used to derive a classical node centrality such as eigenvector centrality, we can define an iterative metric that gradually encodes more global information as the order grows. Then, an iterative metric set that consists of an iterative metric from order 1 to order $K$ is used as input features of the regression model. For example, the number of $k$-hop walks originate from a node, which is determined by the $k$-hop neighborhood of the node, can be derived in an iterative process starting from $k=1$. The resultant iterative metric set is composed of the iterative metric (the number $k$-hop walks) with order $k \in [1, K]$ after $K$ iterations. The benefits of using an iterative metric set to predict nodal influence are as following. Firstly, it allows us to explore to what extent global network information is needed to estimate the nodal influence, i.e., is $K$ necessarily large for accurate prediction? Secondly, It enables us to identify prediction method with low computational complexity, i.e., the regression model with an iterative metric set of a small $K$.  Moreover, in practical applications, one has the flexibility to choose an appropriate $K$ to achieve a well-balanced trade-off between prediction accuracy and computational efficiency. The intuition is illustrated in Figure \ref{fig:illustration}, which shows a network example of $1000$ nodes with community structure generated by Lancichinetti–Fortunato–Radicchi model \cite{lancichinetti2008benchmark}. The red-colored nodes are the top $10\%$ nodes when nodes are ranked by spreading influence (top left), eigenvector centrality\footnote{Eigenvector centrality of a node is the component of the eigenvector corresponds to the largest eigenvalue of the adjacency matrix.} (EC, top middle), degree (DC, top right), number of 2-hop (bottom left), 3-hop (bottom middle) 4-hop (bottom right) walks originating from a node, respectively. The example suggests that the number of $2$-, $3$- and $4$-hop walks possibly reflect nodal spreading influence better than the global metric (eigenvector centrality). Furthermore, it has been observed and partially proved in previous work that a centrality metric like betweenness with a high computational complexity is correlated with local metrics derived from a low order neighborhood \cite{li2015correlation,bartolucci2023ranking}. Hence, global network information, i.e., large $K$, is not necessarily needed in nodal influence prediction.

In this work, we consider three iterative metrics, which converge, respectively to three global node centrality metrics: eigenvector centrality, PageRank centrality \cite{page1999pagerank}, and H index of a node \cite{lu2016h}. The computation of each iterative metric set can be done in $\mathcal{O}(K\cdot |E|)$ time, where $|E|$ is the number of network edges. Based on each iterative metric set, a statistical regression model is built and trained to predict nodal influence. We evaluate the prediction quality of the corresponding three regression models, in comparison with a benchmark \cite{bucur2020top}, i.e., the regression model that uses $7$ nodal centrality metrics, in both real-world networks and synthetic networks with community structure. 
We find that for each iterative metric, the iterative metric set with $K\sim4$ is able to relatively accurately predicts nodal spreading influence, and the prediction quality increases marginally when more global metrics are included as $K$ grows. This suggests the low computational complexity of our iterative metric based prediction methods. Additionally, the best performing iterative metric based model performs comparably with the benchmark model, which has higher computational cost due to the computation of global centrality metrics.

This paper is organized as follows. In Section \ref{sec:methods}{Method}, we introduce the definition of nodal spreading influence and iterative metrics, and regression models to predict nodal influence. Section \ref{sec:results} evaluates the performance of the proposed influence predication methods in both real-world networks and synthetic networks with community structure. Section \ref{sec:discussion} summarizes our findings and discusses limitations and potential extensions of our work.

\section*{Method}
\label{sec:methods}
In this section, we present the definition of nodal spreading influence (Section \ref{ss:SIR}), followed by the definition of iterative metrics (Section \ref{ss:metrics}). We then describe the regression model that uses an iterative metric set to predict nodal spreading influence (Section \ref{ss:regression}).

\subsection*{Nodal spreading influence}
\label{ss:SIR}
We consider the continuous-time Susceptible-Infected-Recovered (SIR) spreading process on a static network \cite{kiss2017mathematics, pastor2015epidemic}. At any time, each node can be in one of three possible states: susceptible, infected, or recovered. At the beginning, one seed node gets infected, while the rest are susceptible. A susceptible node gets infected by each of its infected neighbors at an infection rate $\beta$, and each infected node recovers at a recovery rate $\gamma$. Both the infection and recovery processes are independent Poisson processes. In the steady state, all nodes are either susceptible or recovered. The ratio $\lambda=\beta/\gamma$ is called the effective infection rate. Without loss of generality, we assume recovery rate $\gamma=1$, thus $\lambda=\beta$. For a given network, an epidemic threshold $\lambda_c$ exists. When $\lambda>\lambda_c$, a non-zero fraction of recovered nodes exist in the stable state. When $\lambda<\lambda_c$, the epidemic dies out. The number of recovered nodes in the steady state, or equivalently, the number of nodes that have ever been infected is called the outbreak size. 

The spreading influence of a node is defined as the average outbreak size when the node is chosen as the seed node. We derive the influence of a node as the average outbreak size over $r=10^4$ realizations of the SIR spreading process on a given network.  When the effective infection rate $\lambda\ll\lambda_c$ or when $\lambda \gg \lambda_c$, nodes tend to have similar influence. We focus on predicting influence when the effective infection rate is around the epidemic threshold, e.g., $\lambda=0.5\lambda_c,\lambda_c 1.5\lambda_c, 2\lambda_c$. This is when nodes differ evidently in influence, and influence prediction is crucial. We estimate the epidemic threshold $\lambda_c$ using the numerical approach introduced in \cite{shu2015numerical}. Specifically, referring to $\rho$ as a random variable denoting the influence of a random node in the network, we consider the variability $\sqrt{\langle \rho^2\rangle-\langle \rho\rangle^2}/\langle \rho\rangle$ as a function of $\lambda$. The epidemic threshold $\lambda_c$ is then the value of $\lambda$ that maximizes the variability.
\subsection*{Iterative metrics}
\label{ss:metrics}
Given an undirected network $G=(V, E)$, where $V$ is the set of nodes and $E$ is the set of links between nodes in $V$, the network can be represented by the adjacency matrix $A$, whose element $A_{ij}=1$ if there is a link between node $i$ and $j$, otherwise $A_{ij}=0$. Various node centrality metrics have been proposed to measure topological importance of a node, such as eigenvector centrality, PageRank, and coreness \cite{lu2016h}. For a given a centrality metric, the centralities of all nodes can be denoted by a vector $\mathcal{M}$, where the entry $\mathcal{M}_i$ represents the centrality of node $i$. The iterative process to derive the corresponding iterative metric set starts with an initial metric vector $\mathcal{M}^{(0)}$ and updates the metric vector based on a specific rule $\mathcal{M}^{(k)}=f(\mathcal{M}^{(k-1)})$. Eventually, this process converges to the target centrality metric $\mathcal{M}$. We refer to the derived metric vectors $\{\mathcal{M}^{(k)}, k=1,2,...K\}$ as the  iterative metric set.

In this paper, we consider three iterative processes that converge to three global centrality metrics: eigenvector, PageRank centrality, and coreness of a node, respectively. Three different iterative metrics are derived using these process.
\begin{itemize}
    \item \textbf{Normalized Walk Count (NWC)}. We adopt the power iteration process for the computation of eigenvector centrality to derive the NWC iterative metric. The centrality vector is initialized as the normalized all-one vector  $w^{(0)}=u/\sqrt{N}$, where $u$ is the all-one vector, and is updated iteratively following the updating equation $w^{(k)}=Aw^{(k-1)}/||Aw^{(k-1)}||$. The $k$-th order NWC follows $w^{(k)}=A^{k}u/||A^{k}u||$. Its element $w_i^{(k)}$ represents the normalized number of distinct k-hop walks starting from node $i$ and can be derived from the neighborhood within k hops of the node $i$. As $k$ increases, $w^{(k)}$ converges to the eigenvector centrality $w$. The rate of convergence is determined by the ratio of the largest eigenvalue $\lambda_1(A)$ and the second largest eigenvalue $\lambda_2(A)$ of the adjacency matrix $A$ of the network. The convergence rate is higher when $\frac{|\lambda_2(A)|}{|\lambda_1(A)|}$ is smaller \cite{bjorck2015numerical}. 

    \item \textbf{Visiting Probability (VP)} is derived using the iteration process for the computation of PageRank centrality \cite{page1999pagerank}. The metric vector is initiated as the normalized all-one vector, $p^{(0)}=u/N$, and updated iteratively as $p_i^{(k)}=\alpha \sum_{j=1}^{N}A_{ji}p_j^{(k-1)}/d_j+(1-\alpha)/N$, where $d_j$ is the degree of node $j$ and the teleportation parameter $\alpha$ is set to $0.85$, which is a common choice for calculating the PageRank centrality \cite{gleich2015pagerank}. As $k$ increases, $p_i^{(k)}$ converges to PageRank centrality. The updating equation can be formulated in matrix form: $p^{(k)}=Gp^{(k-1)}$, where $G=\alpha A^TD^{-1}+\frac{1-\alpha}{N}uu^{T}$, matrix $D$ is a diagonal matrix with $D_{ii}=\sum_j A_{ij}$. Since matrix $G$ is a stochastic matrix, the largest eigenvalue $\lambda_1(G)=1$. The rate of convergence is determined by the second largest eigenvalue $\lambda_2(G)$ of the matrix $G$. The smaller $|\lambda_2(G)|$ 
    is, the faster the convergence is \cite{bjorck2015numerical}. The iterative process can be interpreted as a random walk: the walker starts at a randomly selected node. At each time step, with a probability $\alpha$ it moves to a random neighbor of the current visiting node, and with a probability $1-\alpha$ it jumps to a node that is randomly selected from the network. The $k$-th order iterative metric $p_i^{(k)}$ of a node $i$ is the probability that node $i$ is visited by the random walker at the $k$-th hop. Since the information of neighbors' degree is needed in each iteration step, $p_i^{(1)}$ actually encodes $2$-hop neighbors' information. Similarly, the $(k+1)$-hop neighborhood information of a node $i$ is needed to derive $p_i^{(k)}$.
 
    \item \textbf{H index (HI)} \cite{lu2016h}. The $1$-st order H index is defined as the degree of a node, i.e. $HI^{(1)}_i=d_i$. The $k$-th order H index of node $i$ can be derived as $HI^{(k)}_i=\mathcal{H}[HI^{(k-1)}_{j_1}, HI^{(k-1)}_{j_2},...,HI^{(k-1)}_{j_{d_{i}}}]$, where $j_1,...,j_{d_{i}}$ are neighbors of node $i$ and $\mathcal{H}$ is an operator that returns an integer. Specifically, $HI^{(k)}_i$ is the maximum integer such that at least $HI^{(k)}_i$ elements of $[HI^{(k-1)}_{j_1}, HI^{(k-1)}_{j_2},...,HI^{(k-1)}_{j_{d_{i}}}]$ are no less than $HI^{(k)}_i$. It has been proved that $HI^{(k)}$ will converge to the coreness \cite{dorogovtsev2006k, kitsak2010identification} as $k$ increases.
\end{itemize}
The iterative rules $f$ in the three iterative processes only involve operations among a node's 1-hop neighbors. As a result, the metric vector $\mathcal{M}^{(k)}$ after one step iteration encodes information about the neighborhood one hop further than $\mathcal{M}^{(k-1)}$. Given an iterative process, the obtained metric set $\{\mathcal{M}_i^{(1)},\mathcal{M}_i^{(2)},...,\mathcal{M}_i^{(K)}\}$ will be used to predict the influence of node $i$ using the regression model described in Section \ref{ss:regression}. The parameter $K$ controls the scope of information around a node encoded in the iterative metric set $\{\mathcal{M}_i^{(1)},\mathcal{M}_i^{(2)},...,\mathcal{M}_i^{(K)}\}$.
\subsection*{Nodal influence prediction method}
\label{ss:regression}
We assume two key types of information are given to predict nodal influence. Firstly, the network topology is known. Secondly, the influences of a small fraction of nodes are available. In practical scenarios, these influences can often be estimated from real-world diffusion data within social media networks. Our objective is to predict the influences of the remaining nodes in the network. We approach the prediction of nodal influence as a node-level regression problem. Specifically, given a static network $G=(V,E)$ represented by its adjacency matrix $A$ and the spreading influences of a fraction $q$ of nodes, which is randomly selected and denoted as $S_q$, we aim to predict spreading influences of the remaining $1-q$ nodes, referred to as $S_{1-q}$. 

We choose $q=10\%$ assuming only the influences of a small fraction of nodes are known. We train a statistical regression model, which maps the nodal features into the influence of a node, on the training node set $S_q$, and evaluate it on the remaining test node set $S_{1-q}$. For each of the three proposed iterative metrics, the iterative metric set $\{\mathcal{M}_i^{(1)},\mathcal{M}_i^{(2)},...,\mathcal{M}_i^{(K)}\}$ is used as nodal features in the regression model to predict nodal influence.
As a benchmark model, we consider a regression model that uses the same set of $7$ classic centrality metrics as in Bucur's classification model \cite{bucur2020top} as nodal features. These $7$ centrality metrics include both local and global centrality metrics and are able to can complement each other in improving the performance in the node classification task. Finally, we evaluate the prediction quality of the regression models based on $50$ realizations of the random sampling of the training node set $S_q$ and the training of the regression model.

We choose the Random Forest Regression model (RFR), a classic model that captures the nonlinear relationship between input features and the outcome variable, i.e., nodal influence, in our case. We also considered the Ridge regression, a linear regression model with L2 regularization, and obtained similar observations (in Supplementary Information) as the Random Forest Regression.
\section*{Results}
We evaluate the performance of the regression models based on each of the three iterative metrics and the benchmark model based on classic centrality metrics, first in real-world networks in Section \ref{ss:iterativeorder}, and afterwards in synthetic networks with community structures in Section \ref{ss:community}. Finally, we explore the performance of these models in relation to parameters of the spreading process in Section \ref{ss:parameters}.
\label{sec:results}
\subsection*{Networks and measures to evaluate prediction quality}
\label{ss:datasets}

We consider 7 real-world networks that differ in network properties such as size and and diameter (i.e. the largest shortest path length between a node pair among all possible node pairs), including four social networks (advogato, facebook, deezerEU, github), a scientific collaboration networks (Arxiv Astro), a file sharing network (Gnutella04), and an email communication network (Email Enron). We treat all networks as simple, undirected and unweighted. Basic properties of these networks are listed in Table \ref{tab:network_statistics}.
\begin{table}[h!]
\begin{center}
\begin{tabular}{lccccc}
\hline
Dataset  & $|N|$ & $|E|$ & Diameter & $Q$ & $\lambda_c$ \\ \hline
advogato   & 5042    & 41791 & 9 & 0.408 & 0.020 \\
Arxiv-astroph    & 17903   & 196972 & 14 & 0.626 & 0.015 \\
enron      & 33696   & 180811 & 13 & 0.608 & 0.013 \\
facebook   & 63392   & 816886 & 15 & 0.632 & 0.010 \\
gnu04      & 10876   & 39994 & 10 & 0.386 & 0.080 \\
github     & 37700      & 289003    & 11 & 0.453 & 0.011    \\
deezer EU  & 28281      & 92752 &  21 & 0.683 & 0.070 \\
\hline
\end{tabular}
\end{center}
\caption{Basic properties of each real-world network considered: Number of nodes $|N|$, number of edges $|E|$, network diameter, the modularity $Q$ \cite{newman2018networks}, and epidemic threshold $\lambda_C$ of the SIR process on the network.}
\label{tab:network_statistics}
\end{table}

We evaluate the prediction quality of the proposed regression models using the following 3 classic measures:

$R^2$ measures the proportion of the variance in the dependent variable ( nodal influence) that is predictable from the input features in the regression model. $R^2$ is defined as:
\begin{equation}
    R^2=1-\frac{\sum_i(y_i-\hat{y}_i)^2}{\sum_i(y_i-\bar{y})^2}
\end{equation}
Here, $y_i$ and $\hat{y}_i$ are the ground truth and the predicted nodal influence of node $i$ given by the regression model, respectively. $\bar{y}=\frac{1}{n}\sum_{i=1}^{n}y_i$ is the mean value of $y_i$.

\textbf{Kendall's correlation coefficient $\tau(\hat{s}, s)$} measures the similarity of the two ranking lists of nodes based on the predicted nodal influence $\hat{s}$ and the ranking based on the actual nodal influence obtained by SIR simulation. A value of $1$ for $\tau(\hat{s},s)$ indicates that the predicted nodal influence gives the same node ranking as the ground truth, while a value of $-1$ that the two rankings are reverse. Kendall's correlation coefficient \cite{kendall1945treatment} $\tau(\hat{s},s)$ is defined as follows:
\begin{equation}
    \tau(\hat{s}, s) = \frac{n_c-n_d}{\sqrt{(n_c+n_d+T)*(n_c+n_d+U)}}
\end{equation}
where $n_c$ and $n_d$ are the total number of node pairs that are concordant and discordant  \footnote{Node pair $(i,j)$ is concordant if $(\hat{s}_i-\hat{s}_j)(s_i-s_j)>0$, is discordant if $(\hat{s}_i-\hat{s}_j)(s_i-s_j)<0$.} respectively, based on the influence $s$ and the predicted influence $\hat{s}$, $T$ is the number of node pairs that have the same influence but different predicted influence, i.e., $s_i=s_j,\hat{s_i}\neq\hat{s_j}$ and U is the number of node pairs that have the same predicted influence but different influence, i.e., $\hat{s}_i=\hat{s}_j,s_i\neq s_j$.

\textbf{Recognition rate of top-$f\%$} measures the performance of a regression model in identifying the most influential $f\%$ nodes in the test set $S_{1-q}$. It is calculated as the fraction of nodes that are present in the top $f\%$ of both the ranking by predicted nodal influence $\hat{s}$ and the ranking by actual nodal influence $s$. A higher recognition rate of top-$f\%$ implies better performance of the regression model in identifying the most influential nodes.

\subsection*{Performance analysis in real-world networks}
\label{ss:iterativeorder}
We focus on the prediction of spreading influence when the effective infection rate of the SIR spreading process is $\lambda=\lambda_c$, where the epidemic threshold $\lambda_c$ of each network is identified using the method described in Section \ref{ss:SIR}. The values of $\lambda_c$ of each real-world network are shown in Table \ref{tab:network_statistics}. In Section \ref{ss:parameters}, we will discuss how the choice of the effective infection rate around the epidemic threshold impacts on the performance of influence prediction methods.

\begin{figure*}[ht]
\centering
\includegraphics[scale=0.65]{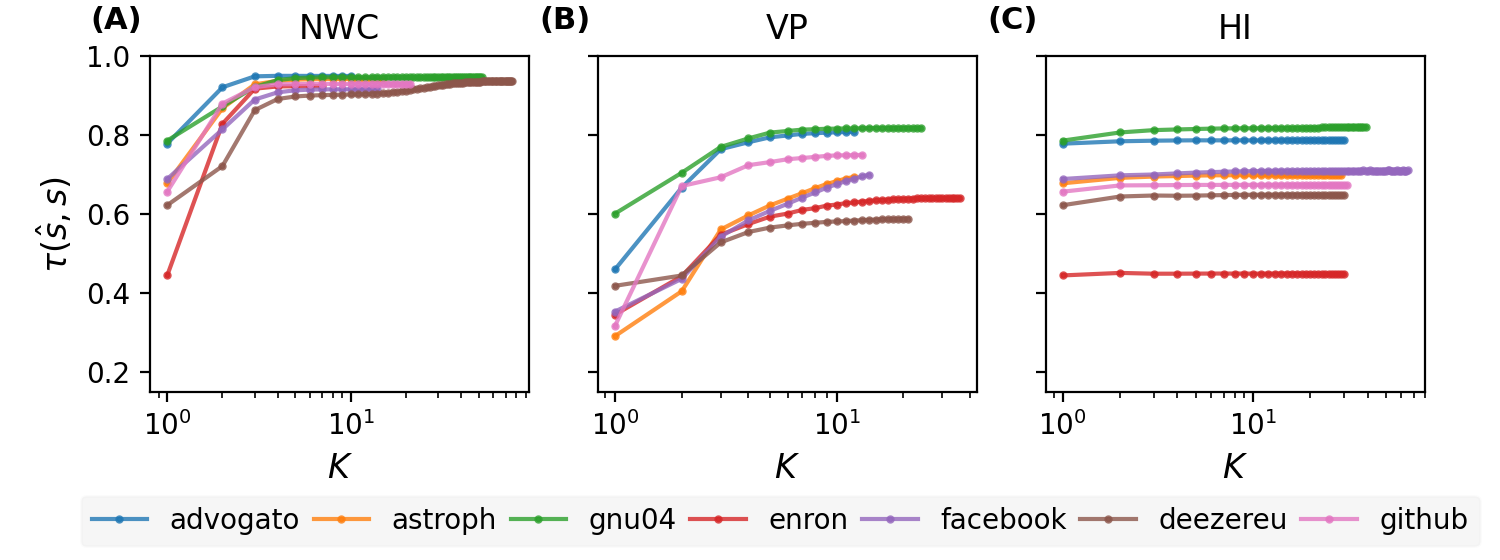}
\caption{Kendall correlation between the actual nodal spreading influence $s$ and the influence $\hat{s}$ predicted by a regression model based on  NWC (panel A), VP (panel B), and H index (panel C) respectively. Results are averaged over $50$ realizations of training set sampling and model training.}
  \label{fig:convergence}
\end{figure*}

We predict nodal influence in real-world networks using the iterative metric based regression models. Each model uses an iterative metric set $\{\mathcal{M}_i^{(1)},\mathcal{M}_i^{(2)},...,\mathcal{M}_i^{(K)}\}$ as input features. Thus, topological information of the $K$-hop ($K+1$-hop for VP) neighborhood of each node is used by the regression model for influence prediction. These regression models are evaluated using the evaluation metrics introduced in Section \ref{ss:datasets}.
In Figure \ref{fig:convergence}, we show the Kendall correlations $\tau(\hat{s},s)$ between the actual nodal influence $s$ and the influence $\hat{s}$ predicted by a regression model as a function of $K$ in real-world networks. As $K$ grows, higher order iterative metrics are included, and the prediction quality increases. Similar trends are observed for other evaluation metrics of prediction quality (see Supplementary Information).

\begin{figure*}[ht]
\centering
\includegraphics[width=\textwidth]{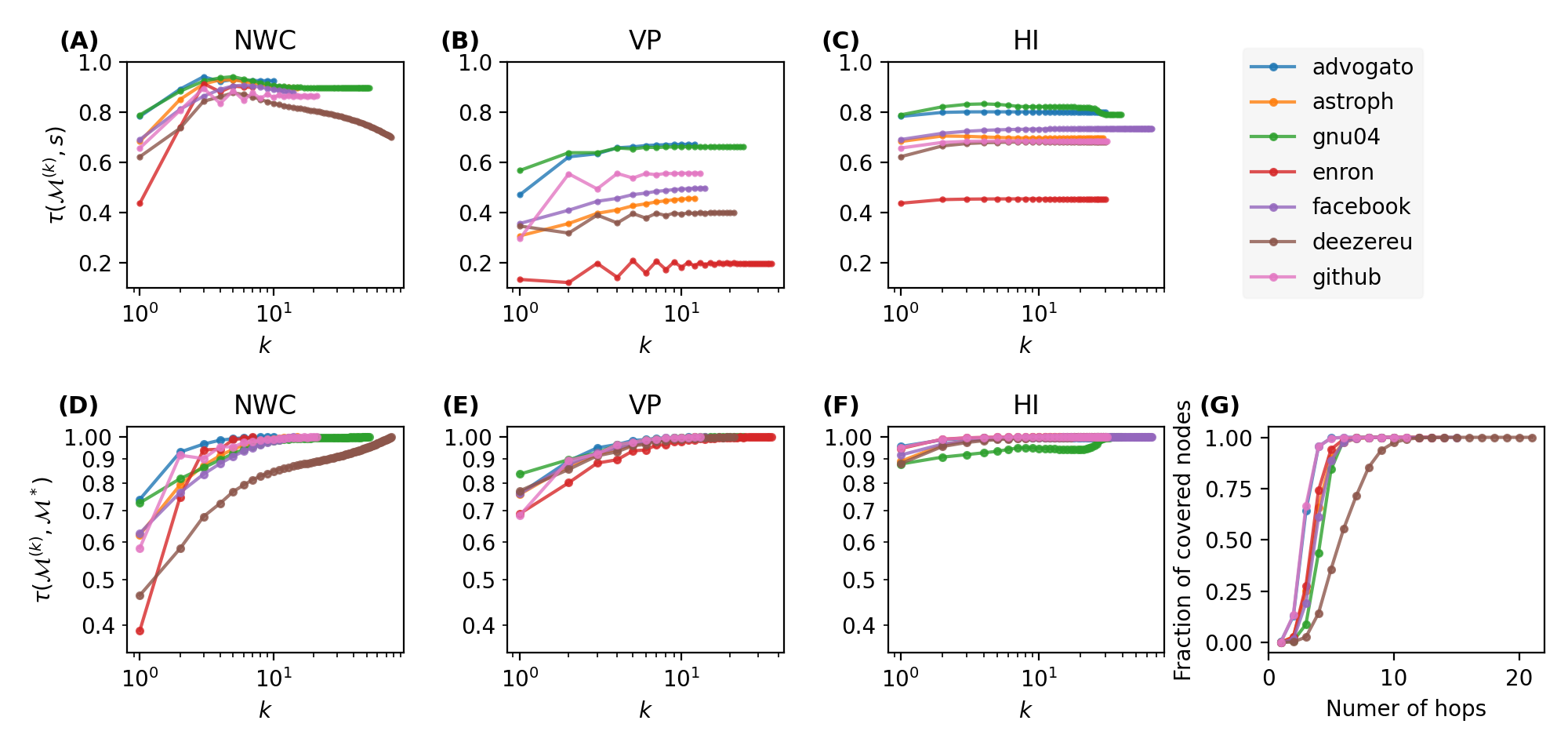}
\caption{Kendall correlation between nodal spreading influence $s$ and different orders of NWC ($w^{(k)}$, panel A), VP ($p^{(k)}$, panel B), and H index ($h^{(j)}$, panel C), and the convergence of NWC (D), VP (E), HI (F), measured by the Kendall's correlation between the iterative metric after $k$ iterations and the corresponding global centrality metrics, as a function of iteration number $k$ in $7$ real-world networks. (G) shows the coverage, i.e. the average fraction of nodes covered by hopping step out from a node, as a function of the number of hops.}
\label{fig:convergence_metrics}
\end{figure*}

A notable observation in Figure \ref{fig:convergence} is that, for all three iterative metrics, the prediction quality is already close to the highest when $K\sim4$ and only increases marginally by choosing a $K>4$. This suggests that a regressing model using relatively local topological information could already achieve comparably good predication quality as that using more global information.
To understand this, we first explore the correlation $\tau(\mathcal{M}^{(k)}, s)$ between the $k$-th order iterative metric $\mathcal{M}^{(k)}$ and the spreading influence $s$. The correlation is shown in Figure \ref{fig:convergence_metrics} (A-C) for each iterative metric, respectively. Generally, all three metrics at any order $k$ exhibit positive correlation with spreading influence, which indicates that each iterative metric has certain predictive power. As $k$ increases, the correlation $\tau(\mathcal{M}^{(k)}, s)$ increases when $k$ is small, and achieves (nearly) the highest around $k\sim4$ in all considered real-world networks, after which the correlation increases slightly (or declines in case of NWC). This suggests that, in addition to local topological properties, semi-local centrality metrics are likely needed for nodal influence prediction and high order iterative metrics that encode global topological information is probably unnecessary. 

Secondly, we study the convergence of each iterative metric itself. As $k$ increases, each $\mathcal{M}^{(k)}$ centrality metric converges to the global centrality metric $\mathcal{M}^{*}$. 
According to the definition, the three iterative metrics converges to three global metrics: eigenvector centrality, PageRank centrality, and coreness respectively. Figure \ref{fig:convergence_metrics} (D-F) shows the Kendall's correlation $\tau(\mathcal{M}^{(k)}, \mathcal{M}^*)$ between the $k$-th order metric $\mathcal{M}^{(k)}$ and the global metric $\mathcal{M}^{*}$ as a function of $k$ for each iterative metric. 
For each iterative metric, $\mathcal{M}^{(k)}$ converges to $\mathcal{M}^{*}$ with different convergence rates in different networks. 
Importantly, $\mathcal{M}^{(k)}$ exhibits relatively high correlation with $\mathcal{M}^{*}$ at $k\sim4$ in most networks. This finding partly explains why the corresponding regression model improves in prediction quality only marginally as $K$ increases when $K\geq4$.
Furthermore, the large correlation $\tau(h^{(k)}, h^*)$ for any $k$, as shown in Figure \ref{fig:convergence_metrics} (F) explains why the prediction quality of the regression model based on HI hardly improves when $K$ grows, as observed in Figure \ref{fig:convergence} (C). Finally, we noted that network \textit{deezer EU} has a relatively lower Kendall's correlation $\tau(\mathcal{M}^{(k)}, \mathcal{M}^*)$ in case of NWC than the other networks, as shown in Figure \ref{fig:convergence_metrics} (D). This is likely due to its high modularity (shown in Table \ref{tab:network_statistics}), motivating us to investigate the impact of the strength of community structure on nodal influence prediction in the next section.

To gain insight into why each iterative metric $\mathcal{M}^{(k)}$ exhibits relatively high correlation with $\mathcal{M}^{*}$ at $k\sim4$ in most networks, we investigate the average size of the $k$-hop neighborhood, i.e., the fraction of nodes that is reachable (covered) from a random node in $k$ hops. Figure \ref{fig:convergence_metrics} (G) shows that in most real-world networks, a significant fraction of nodes is reachable from a random node within $4$ hops. Hence, an order $4$ iterative metric captures the topological information of a significant amount of nodes, supporting why $\tau(\mathcal{M}^{(k)}, \mathcal{M}^*)$ is high when $k\sim4$. Network \textit{deezer EU} differs from the other real-world networks: its $4$-hop coverage is lower and network diameter is larger as shown in Table \ref{tab:network_statistics}, which is likely due to its community structure. Correspondingly, $\tau(\mathcal{M}^{(k)}, \mathcal{M}^*)$ when $k\sim4$ for NMC is lower in \textit{deezer EU} than in the other networks. 

Among all three iterative metrics, NWC achieves evidently the highest prediction quality when $K\geq4$. This can be explained by the higher correlation $\tau(s,w^{(k)})$ between the NWC centrality $w^{(k)}$ and the spreading influence $s$ at each order $k$, as shown in Figure \ref{fig:convergence_metrics} (A-C).

It has been found that combining local and global node centrality metrics can more accurately identify top influencers than using either local or global centralities alone \cite{bucur2020top}. Hence, we build a benchmark regression model that uses the same $7$ centrality metrics (local ones, e.g., degree, and global ones, e.g., betweenness) as in the classification model in \cite{bucur2020top} as input features. Now, we compare the prediction quality of the proposed iterative metric based models with the benchmark model. We choose $K=4$ for iterative metric based models to ensure computational efficiency and reasonably good predication quality.

\begin{figure*}[ht]
\centering
\includegraphics[width=\textwidth]{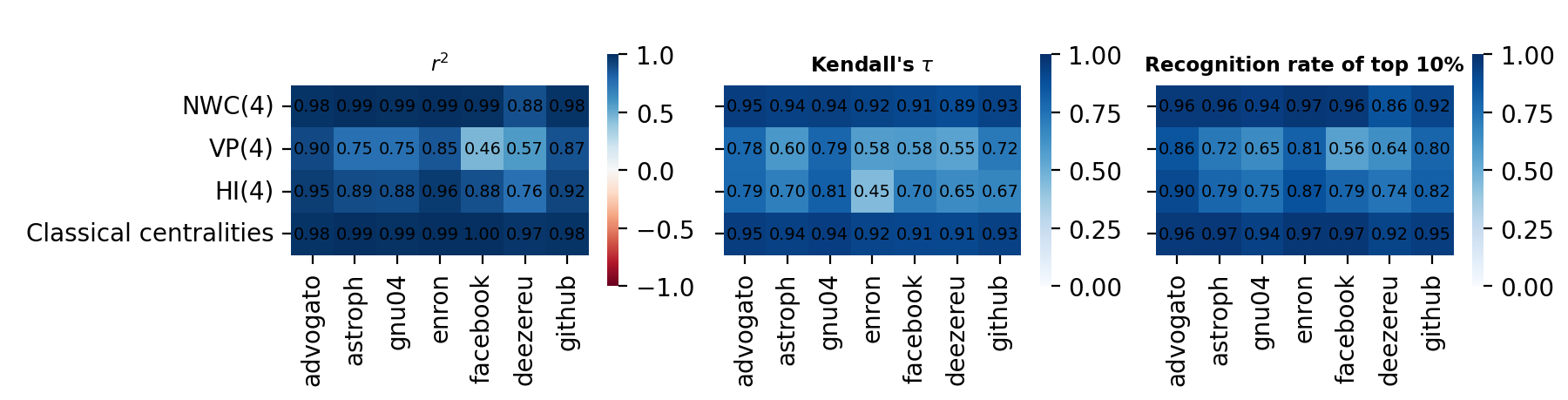}
\caption{Prediction performance comparison across different empirical networks (horizontal axis) of four sets of metrics (vertical axis): Normalized Walk Count when $K=4$ (NWC(4)), Visiting Probability  when $K=4$ (VP(4)), H index  when $K=4$ (HI(4)), and seven centralities \cite{bucur2020top}. Three panels correspond to different evaluation measures of predictive models: $r^2$ (left panel), Kendall's $\tau$ (middle panel), and recognition rate of top $10\%$ nodes (right panel), respectively. Results are averaged over $50$ realizations of Random Forest Model training process.}
  \label{fig:comparison}
\end{figure*}

Figure \ref{fig:comparison} shows three evaluation measures of the regression models: $r^2$ (left panel), Kendall correlation between the actual nodal spreading influence $s$ and the predicted influence $\hat{s}$ of the node by a regression model (middle panel), and the recognition rate of top $10\%$ nodes (right panel). Across all real-world networks, we find that NWC based model and the benchmark model achieve comparable prediction quality and perform significantly better than the other two iterative metric based models.

Moreover, the computational complexity of NWC based model is lower than that of the benchmark model, which requires the computation of global centrality metrics. For example, the computational complexity of betweenness and closeness is $\mathcal{O}(|V|\cdot |E|)$, where $|V|$ is the number of nodes and $|E|$ is the number of edges in the network. The computation of an iterative metric engages the operation with only $1$-hop neighbors in each iteration. The computational complexity of the iterative metric set $\{\mathcal{M}_i^{(1)},\mathcal{M}_i^{(2)},...,\mathcal{M}_i^{(K)}\}$ for all nodes equals that of $\mathcal{M}_i^{(K)}$ for all nodes, which is $\mathcal{O}(K\cdot |E|)$, and a relative small $K$ facilitates its application in large-scale networks.

\subsection*{Prediction on networks with communities}
\label{ss:community}
Community structure has been observed in many real-world networks \cite{fortunato2016community}, where nodes within a community are densely connected while nodes from different communities have fewer connections. The existence of communities affects significantly the spreading process unfolding on a network \cite{saxena2018social, kumar2018efficient} and has been ignored in most centrality metrics used to predict nodal influence \cite{rajeh2021characterizing, costantini2022measuring}. Here we evaluate the performance of our influence prediction methods in networks with community structures and investigate how community structure affects the prediction quality.  
To this aim, we adopt the Lancichinetti–Fortunato–Radicchi (LFR) model \cite{lancichinetti2008benchmark} to generate networks with power-law degree distribution and community size distribution, as observed in real-world networks. One advantage of LFR model is that the strength of the community structure in the generated networks can be changed via tuning its parameters. We use LFR model to generate networks with the following properties: network size $N=1000$, the exponent of the power-law degree distribution $\tau_1=2$, and exponent of the power-law community size distribution $\tau_2=3$, the average degree $\langle k\rangle=10$, the maximum degree $k_{max}=50$, the range of community sizes $[50,100]$. The mixing parameter $\mu$ represents the fraction of inter-community links of a node. When $\mu=0$, the generated networks have the strongest community structure, with communities being disjoint from each other. The model with $\mu=1$ generates networks where all links fall between different clusters. When $\mu>0.5$, the community structure is not evident anymore \cite{lancichinetti2008benchmark}. We set $\mu=[0.02, 0.05, 0.1, 0.2, 0.3, 0.4]$, thus six networks with different strength of communities are generated. Properties of these generated networks are listed in Table \ref{tab:network_statistics_lfr}.
\begin{table}[!h]
    \centering
    \begin{tabular}{rccc}
    \hline
        $\mu$ & Diameter & $Q$ & $\lambda_c$ \\ \hline
        0.02 & 10 & 0.924 & 0.090 \\ \hline
        0.05 & 6 & 0.872 & 0.080 \\ \hline
        0.1 & 5 & 0.608 & 0.070 \\ \hline
        0.2 & 5 & 0.632 & 0.070 \\ \hline
        0.3 & 5 & 0.386 & 0.070 \\ \hline
        0.4 & 5 & 0.453 & 0.070 \\ \hline
    \end{tabular}
\caption{Basic properties of networks generated by LFR model using different mixing parameter $\mu$: network diameter, the modularity $Q$, epidemic threshold $\lambda_C$ of the SIR process on the network.}
\label{tab:network_statistics_lfr}
\end{table}

\begin{figure*}[ht]
\centering
\includegraphics[scale=0.65]{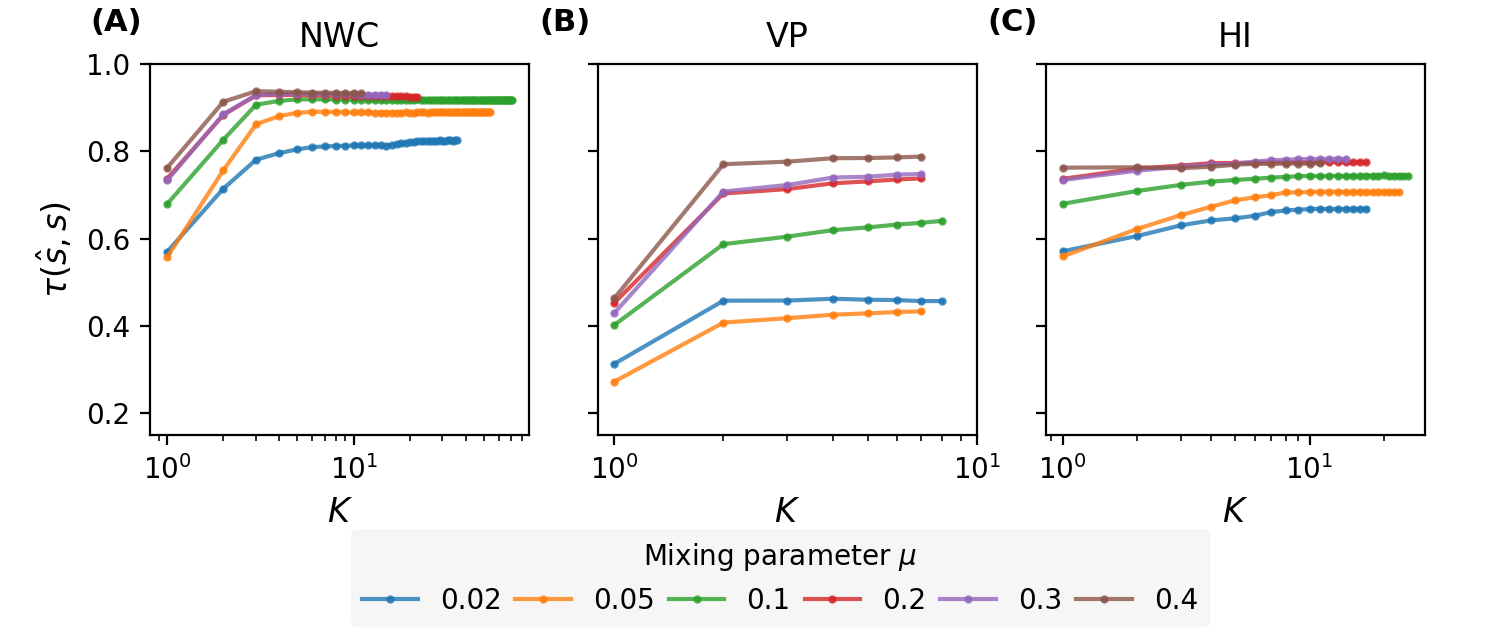}
\caption{Kendall correlation between nodal spreading influence $\hat{s}$ predicted by different numbers of iterative metrics as features and nodal spreading influence given by SIR simulations of NWC (A), VP (B), and H index (C). Results are averaged over $50$ realizations of training set sampling and model training.}
  \label{fig:convergence_LFR}
\end{figure*}

\begin{figure*}[ht]
\centering
\includegraphics[width=\textwidth]{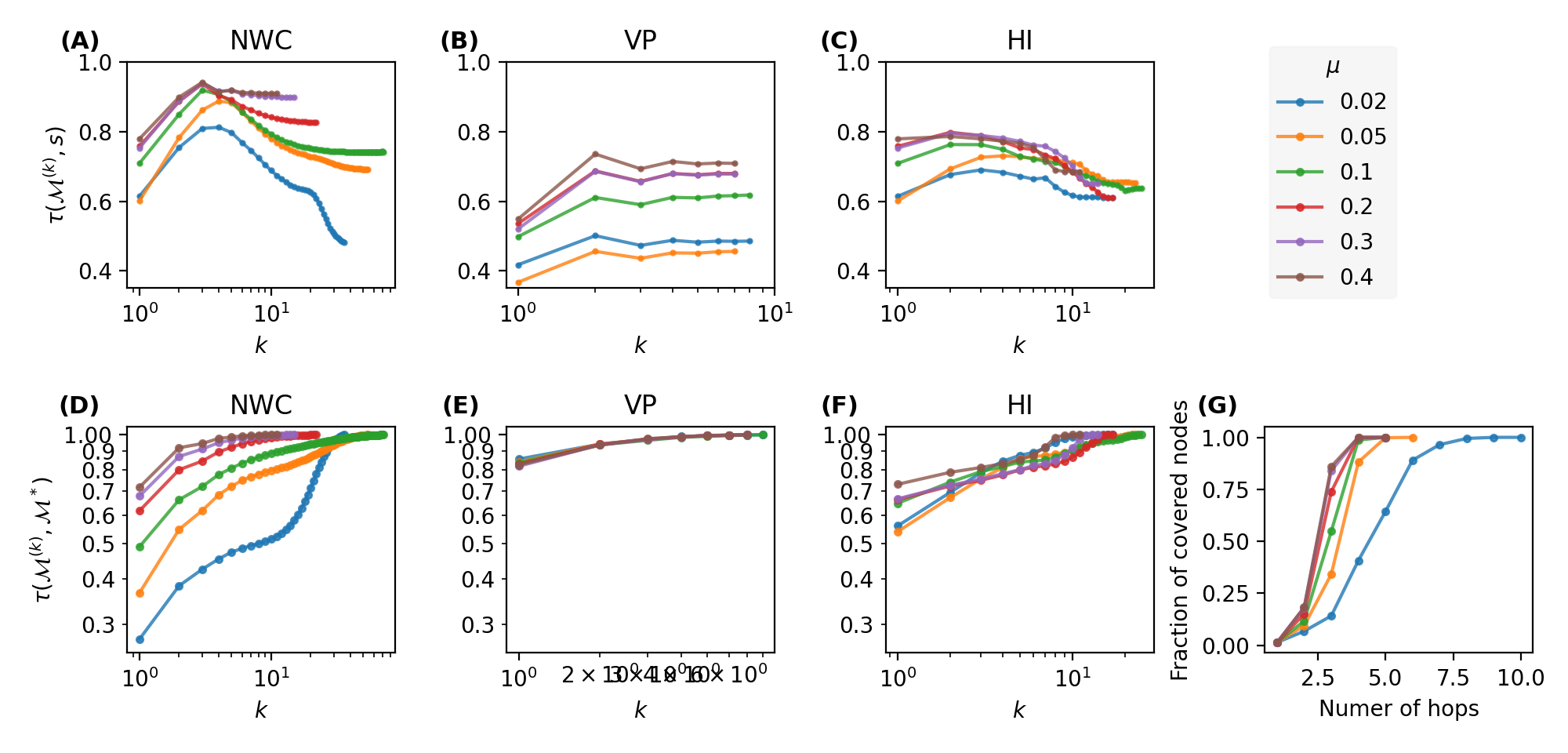}
\caption{Kendall correlation between nodal spreading influence $s$ and different orders of NWC ($w^{(k)}$, panel A), VP ($p^{(k)}$, panel B), and H index ($h^{(j)}$, panel C), and the convergence of NWC (D), VP (E), HI (F), measured by the Kendall's correlation between the iterative metric after $k$ iterations and the corresponding global centrality metrics, as a function of iteration number $k$ in Lancichinetti–Fortunato–Radicchi (LFR) networks with different $\mu=0.02, 0.05, 0.1, 0.2, 0.3, 0.4$. (G) shows the coverage, i.e. the average fraction of nodes covered by hopping step out from a node, as a function of the number of hops.}
  \label{fig:convergence_metrics_LFR}
\end{figure*}

We first evaluate our iterative metric based models in predicting nodal influence in LFR networks when the effective infection rate of the SIR model is around epidemic threshold, i.e., $\lambda=\lambda_c$. Figure \ref{fig:convergence_LFR} (A-C) show Kendall correlations $\tau(\hat{s}, s)$ between the nodal spreading influence $s$ and the prediction $\hat{s}$ by a regression model based on an iterative metric set $\{\mathcal{M}^{(1)},\mathcal{M}^{(2)},...,\mathcal{M}^{(K)}\}$, as a function of $K$ in LFR networks. 
Like what we observed in real-world networks, the prediction quality increases as $k$ increases since more nodal information (features) are included. Notably, the prediction quality only improves marginally when choosing a $K>4$. This can be understood by the correlation $\tau(\mathcal{M}^{(k)}, s)$ between $\mathcal{M}^{(k)}$ and nodal influence $s$, which is shown in Figure \ref{fig:convergence_metrics_LFR} (A-C). As $k$ increases up to $k\sim4$, the correlation $\tau(\mathcal{M}^{(k)}, s)$ increases. As $k$ increases further, the correlation tends to decrease, which differs from what we have observed in real-world networks. This suggests that high-order ($k>4$) iterative metrics are less predictive than an iterative metric of an order around $k=4$, thus less needed to predict nodal influence. Furthermore, we explore the convergence of an iterative metric $\mathcal{M}^{(k)}$ as $k$ increases. Figure \ref{fig:convergence_metrics_LFR} (D-F) show the Kendall's correlation $\tau(\mathcal{M}^{(k)}, \mathcal{M}^*)$ as a function of $k$ for the three iterative metrics, respectively. 
For NWC, the correlation tends to be lower when $k\sim4$ as the mixing parameter $\mu$ gets smaller or equivalently in network with more evident community structure. Still, the prediction quality of the regression models is close to optimal when $K\sim4$, since the higher order metric is less predictive. This is also in line with the intuition that in networks with strong community structure and when the infection rate is around the critical epidemic threshold, nodal influence is supposed to be mainly determined by nodal property derived within or around  the community that the node belongs to. 

Figures \ref{fig:convergence_metrics_LFR} (G) shows the average fraction of nodes that are reachable (covered) from a randomly chosen node within $k$ hops neighborhood, i.e., the so called coverage, as a function of $k$. In networks with strong community structure (small $\mu$), the coverage and $\tau(\mathcal{M}^{(k)}, \mathcal{M}^*)$ when $k\sim4$ tend to be small. In such networks, an order $k\sim4$ iterative metric encodes topological information of a small fraction of nodes, which explains partially the weak correlation $\tau(\mathcal{M}^{(k)}, \mathcal{M}^*)$ when $k\sim4$.

\begin{figure*}[ht]
\centering
\includegraphics[width=\textwidth]{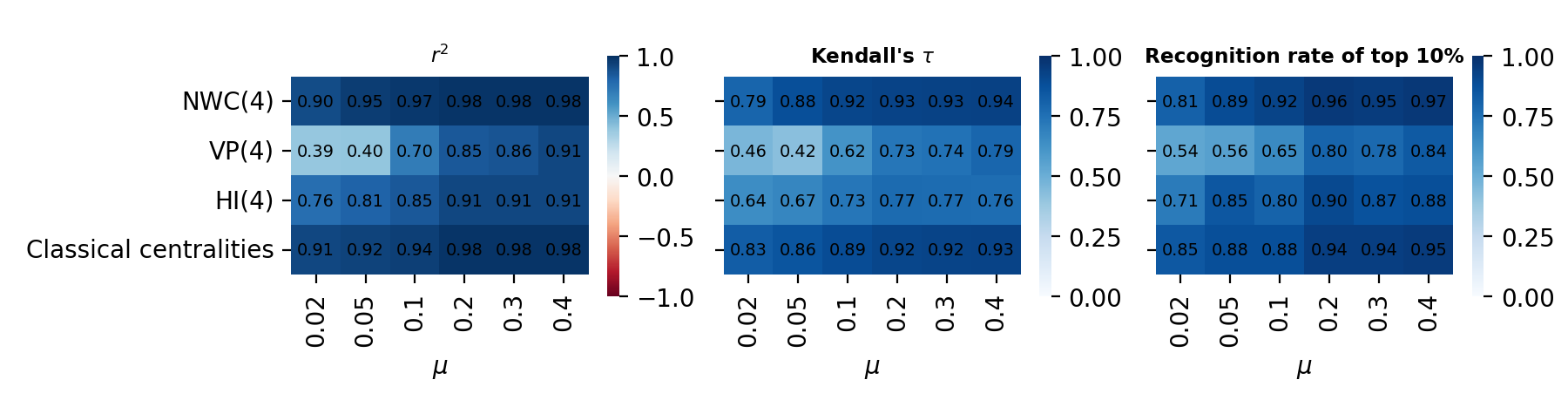}
\caption{Prediction performance on model networks generated with LFR model with varying mixing parameter $\mu$ (horizontal axis) of five sets of metrics (vertical axis): Normalized Walk Count when $K=4$ (NWC(4)), Visiting Probability when $K=4$ (VP(4)), H index when $K=4$ (HI(4)), and classical centralities. Three panels correspond to different evaluation measures of predictive models: $r^2$ (left panel), kendall's $\tau$ (middle panel), and recognition rate of top $10\%$ nodes (right panel), respectively. Results are averaged over $50$ realizations of training process of Random Forest Model.}
  \label{fig:LFR_comparison}
\end{figure*}

Now we compare the prediction quality of iterative metric based models (when $K=4$) with the benchmark model in LFR networks via the same three evaluation measures as in real-world networks. Figure \ref{fig:LFR_comparison}, shows that NWC based model and the benchmark model are comparably the most predictive and HI performs better than VP. As the strength of community grows, all models perform worse. This can be explained by the small (large) correlation $\tau(\mathcal{M}^{(k)}, s)$ in networks with a strong (weak) community structure, as shown in Figure \ref{fig:convergence_metrics_LFR} (A-C). 

\subsection*{Prediction of nodal spreading influence near epidemic threshold}
\label{ss:parameters}
\begin{figure*}[ht]
\centering
\includegraphics[scale=0.75]{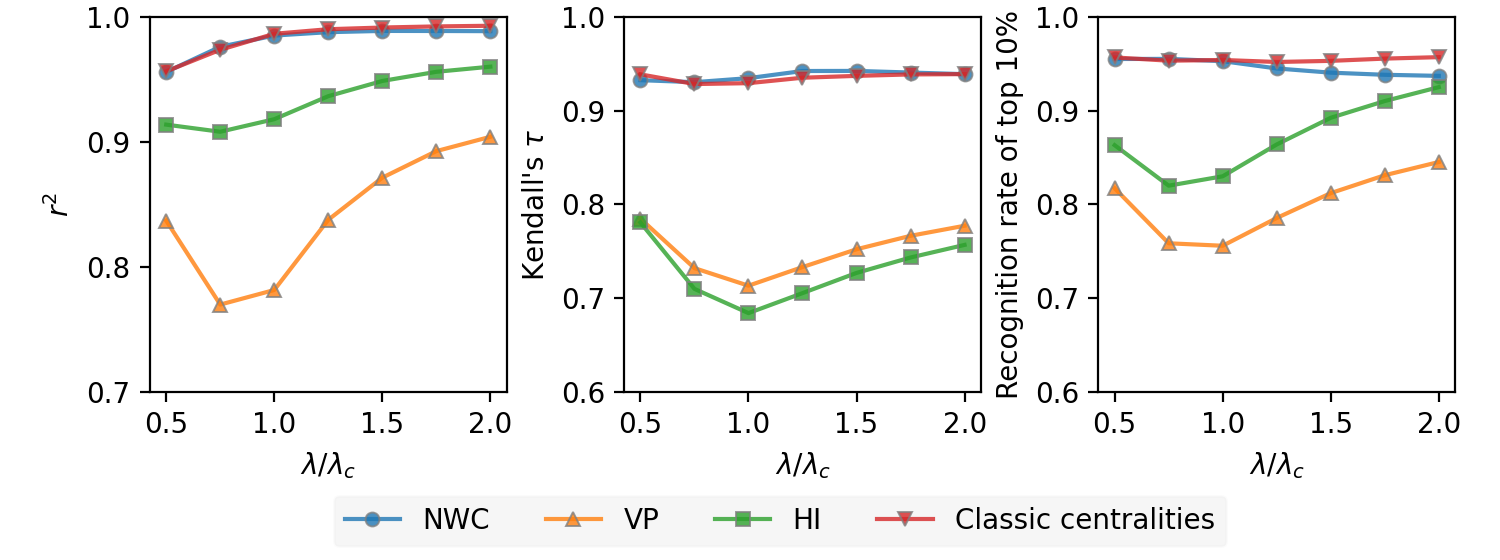}
\caption{Average prediction quality over $7$ real-world networks (shown in Table \ref{tab:network_statistics}) as a function of $\lambda/\lambda_c$ of $4$ different metric sets: Normalized Walk Counts (NWC), Visiting Probability (VP), H index (HI), and $7$ node centralities \cite{bucur2020top}. Three panels correspond to different evaluation measures of predictive models: $r^2$ (left panel), kendall's $\tau$ (middle panel), and recognition rate of top $10\%$ nodes (right panel), respectively. Results are averaged over $50$ realizations of training process of Random Forest Model.}
  \label{fig:averageall}
\end{figure*}

So far, we have focus on the influence prediction problem, where the influence is defined for the SIR epidemic spreading process with $\lambda=\lambda_c$. It has been shown that the change of parameters in the epidemic spreading can lead to different rankings of nodes according to their influences \cite{vsikic2013epidemic,liu2016locating, qu2017ranking}. Hence, we evaluate the prediction quality of all the models when the effective infection rate varies around the epidemic threshold $\lambda_c$ in real-world networks. Figure \ref{fig:averageall} shows that, as $\lambda$ varies from $0.5\cdot\lambda_c$ to $2.0\cdot\lambda_c$, NWC and the benchmark model show comparable prediction quality, which is better than HI and VP and less sensitive to $\lambda$. Compared with HI, VP achieves slightly better Kendall's rank correlation $\tau$, while HI shows significantly large recognition rate of top $10\%$ nodes. In summary, NWC based model using relatively local topological information of a node, performs comparably well as the benchmark model but has low computational complexity, as evaluated across different networks and effective infection rates around the epidemic threshold.

\section*{Discussion and future work}
\label{sec:discussion}

In summary, we explore to what extent local and global topological information of a node is needed for the prediction of nodal spreading influence and whether relatively local topological information around a node is sufficient for the prediction. We propose to predict nodal influence by an iterative metric set derived from an iterative process. Three iterative metrics are considered: Normalized Walk Counts (NWC), Visiting Probability (VP), and H index (HI), which converge to eigenvector centrality, PageRank, and H index, respectively. The regression model using an iterative metric set as input features is trained on a fraction of nodes whose influence is known and is used to predict the nodal influence of the remaining nodes. We evaluate and interpret the performance of these three iterative metric based models in predicting nodal influence in SIR spreading processes with diverse effective infection rates around the epidemic threshold, on both real-world networks and synthetic networks with different strength of community structure. We find that, an iterative metric set including an iterative metric of relatively low orders, i.e., up to order $K\sim4$, could achieve comparable prediction quality to those incorporating iterative metric of orders up to $K>4$. The addition of a high-order iterative metric that encloses global topological information around a node improves the prediction quality only marginally. This can be understood by the observation that the correlation between an iterative metric of order $k$ and nodal influence approaches the maximum when $k\sim4$ and the relatively fast convergence of each iterative metric. We compare the prediction quality of these three iterative metric based models when $K=4$ with the benchmark model that uses $7$ classic local and global centrality metrics. It has been found that the iterative metric NWC based model achieves comparable prediction quality with the benchmark model, while VP and HI based models exhibit lower predictive power. This suggests that the NWC metric of relatively low orders contain sufficient information to predict nodal influence reasonably well. Correspondingly, the computation complexity of NWC based model is lower than that of the benchmark model.  

In this work, we confine ourselves to the influence of nodes in the SIR spreading process on a static network. In many cases, epidemics and information spread via the time-evolving networks, which are even possibly high-order temporal networks \cite{ceria2023temporal, cencetti2021temporal}. Our proposed method can be extended to explore possibility of predicting nodal influence defined in such more complex context using local network information. 
\section*{Acknowledgment}

This publication is supported by the project FORT-PORT (with project number KICH1.VE03.21.008 of the research programme KIC - MISSION 2021 which is (partly) financed by the Dutch Research Council (NWO).
\bibliography{sample}







\end{document}